\documentclass[aps,onecolumn,showpacs,nofootinbib,showkeys]{revtex4-2}
\usepackage[margin=3.15cm]{geometry}

\RequirePackage[T1]{fontenc}

\usepackage{epsfig,graphicx,amsmath,amssymb,bm}
\RequirePackage{mathptmx}      
\usepackage{mathrsfs}
\usepackage{amssymb}
\RequirePackage{color}
\usepackage{changes}
\usepackage{slashed}
\usepackage{multirow}
\usepackage{scalerel}
\usepackage{tikz-feynman}
\usepackage{textcomp}
\usepackage{subcaption}
\usepackage[english]{babel}
\usepackage{float}
\RequirePackage{hyperref}
\hypersetup{
    linktocpage,
    colorlinks,
    citecolor=blue,
    filecolor=black,
    linkcolor=blue,
    urlcolor=blue,
}
\usepackage{nccmath}

\begin{document}

\title{Semileptonic decays of $\rho$, $\omega$, $\phi$, $\eta$ and $\eta'$ mesons in ground and first radially excited states in the $U(3)\times U(3)$ NJL model}


\author{Mikhail K. Volkov$^{1}$}\email{volkov@theor.jinr.ru}
\author{Kanat Nurlan$^{1,2,3}$}\email{nurlan@theor.jinr.ru}

\affiliation{$^1$ Bogoliubov Laboratory of Theoretical Physics, JINR, 
                 141980 Dubna, Moscow region, Russia \\
                $^2$ The Institute of Nuclear Physics, Almaty, 050032, Kazakhstan\\
                $^3$ Al-Farabi Kazakh National University, Almaty, 050040 Kazakhstan}   


\begin{abstract}
In the extended $U(3)\times U(3)$ NJL model, semileptonic decays of $\rho$, $\omega$, $\phi$, $\eta$ and $\eta'$ mesons in both the ground and first radially excited states are described. Quark loops of the anomalous type are considered, which are based on the vertex $\eta \to VV$. The obtained results are in satisfactory agreement with the existing experimental data. In addition, a number of predictions for experimentally unmeasured decays of radially excited mesons are presented.


\end{abstract}

\pacs{}

\maketitle


\section{\label{Intro}Introduction}
Studies of meson interactions at low energies are of great interest for investigations of internal structures and interactions of mesons with each other. Unfortunately, for a description of meson interactions in the low-energy region, it is impossible to use the perturbation theory of the fundamental QCD theory due to a large value of the coupling constant. Therefore, one has to use various phenomenological theories based on internal symmetries of strong interactions. The most frequently used models in specific calculations are the Chiral Perturbation Theory \cite{Gasser:1983yg, Gasser:1984gg} and Nambu-Jona-Lasinio type chiral symmetric models \cite{Nambu:1961tp, Eguchi:1976iz, Ebert:1982pk, Volkov:1984kq, Volkov:1986zb, Ebert:1985kz, Vogl:1991qt, Klevansky:1992qe, Volkov:1993jw, Hatsuda:1994pi, Ebert:1994mf, Buballa:2003qv, Volkov:2005kw, Volkov:2017arr, Volkov:2022jfr}. 

Among the mesons of a pseudoscalar nonet described in the $U(3)\times U(3)$ NJL model, $\eta$ and $\eta'$ mesons are of special interest \cite{Volkov:2022jfr, Gan:2020aco}. Indeed, unlike pions and kaons, singlet-octet mixing occurs in $\eta$ and $\eta'$ mesons. As a result, these mesons contain quarks of three flavors. The mixing of light $u$ and $d$ quarks with a heavier $s$ quark occurs due to the gluon anomaly (the interaction of gluons with quarks). In the NJL model, this effect is quite satisfactorily described using the 't Hooft interaction \cite{tHooft:1976rip, Volkov:1998ax}.  

In the present paper, we are interested in the decays of $\rho$, $\omega$, $\phi$, $\eta$ and $\eta'$ mesons, which are described through triangular quark diagrams of an anomalous type with the production of a lepton pair $l^+l^-$ in the final state. We consider $\rho$, $\omega$, $\phi$, $\eta$ and $\eta'$ mesons in both the ground and first radially excited states. In our work, when describing the production of a lepton pair, we proceed from the existence of lepton universality. Namely, the probabilities of the lepton pairs $e^+e^-$ and $\mu^+\mu^-$ production are described by the same coupling constant and only differ in masses. 

In addition to numerous theoretical calculations, great interest in the description of $\eta$ and $\eta'$ mesons is also manifested in experimental physics. Recently, the BESIII collaboration presented new data on the $\eta' \to \rho e^+e^-$, $\eta' \to \omega e^+e^-$ and $\eta' \to \pi^+\pi^- \mu^+\mu^-$ decays obtained at the electron-positron collider \cite{BESIII:2013tjj, BESIII:2015jiz, BESIII:2020elh}. The KLOE experiment at the $\phi$-factory DA$\Phi$NE with high statistics improved the branching fractions of $\phi \to \eta e^+e^-$\cite{KLOE-2:2011hhj, KLOE-2:2014hfk}. The production of $\eta$ mesons in proton-proton collisions and in the $pd \to^3\text{He} \eta$ reaction were studied at the COSY synchrotron storage ring \cite{Smyrski:2007nu, Xie:2016zhs, Husken:2019dou}. Also, in the Crystal Ball at MAMI experiment, the internal properties of eta mesons were determined by measuring the reaction threshold $\gamma p \to \eta p$ \cite{Unverzagt:2009vm}.

It is important to note that a program is being developed for a new experimental $\eta$ meson factory REDTOP \cite{Gatto:2019dhj, REDTOP:2022slw}, where mesons will be generated by a proton beam at Fermilab or at the proton synchrotron at CERN. It is expected that the number of events at the new factory will be 2 orders of magnitude greater than previous experiments. In the planned experiments, in particular, semileptonic decays with high statistics will be studied. Also, high statistics are expected in the planned experiments at the Super Charm-Tau Factories, where the corresponding decay widths can be measured with good accuracy \cite{Charm-TauFactory:2013cnj, Luo:2018njj}. 

All the noted experiments show the relevance of the study of $\eta$ meson physics. Therefore, it seems interesting to give a theoretical description of semileptonic decays, especially involving $\eta$ and $\eta'$ mesons in the first radially excited states, including a number of predictions for future experiments. 

\section{Quark-meson Lagrangians of the NJL model}
The $U(3) \times U(3)$ chirally symmetric NJL model will be used to describe semileptonic decays. The quark–meson Lagrangian for the strong interaction of $\rho$, $\omega$, $\phi$, $\eta$ and $\eta'$ mesons in the ground and first radially excited states in the NJL model takes the form \cite{Volkov:2017arr, Volkov:2022jfr, Volkov:1996br, Volkov:1996fk}
\begin{eqnarray}
{\cal L}_{V+V'} = \bar{q} \biggl[\frac{1}{2} \gamma^{\mu}\lambda_{V}(a_{V}V_{\mu} + b_{V}V'_{\mu})
\biggl]q,
\end{eqnarray}

        \begin{eqnarray}
            \mathcal{L}_{\eta} = \bar{q} i\gamma^{5} \sum_{i = u, s} \lambda_{i} \left[A^{i}_{\eta}\eta + A^{i}_{\eta'}\eta' + A^{i}_{\hat{\eta}}\hat{\eta} + A^{i}_{\hat{\eta}'}\hat{\eta}'\right]q,
        \end{eqnarray}
where $V=\rho, \omega, \phi$ and $V'=\rho', \omega', \phi'$; $q$ and $\bar{q}$ are u, d and s quark fields with constituent quark masses 
$m_{u} \approx m_{d} = 270$~MeV, $m_{s} = 420$~MeV; excited mesonic states of $\eta$ and $\eta'$ mesons are 
marked with a hat, $\lambda$ are linear combinations of the Gell-Mann matrices 
\cite{Volkov:2017arr},
\begin{eqnarray}
\label{verteces1}
	a_{V} = \frac{1}{\sin(2\theta_{M}^{0})}\left[g_{M}\sin(\theta_{M} + \theta_{M}^{0}) +
	g'_{M}f_{M}(k_{\perp}^{2})\sin(\theta_{M} - \theta_{M}^{0})\right], \nonumber\\
	b_{V} = \frac{-1}{\sin(2\theta_{M}^{0})}\left[g_{M}\cos(\theta_{M} + \theta_{M}^{0}) +
	g'_{M}f_{M}(k_{\perp}^{2})\cos(\theta_{M} - \theta_{M}^{0})\right],
\end{eqnarray}
        
The subscript M indicates the corresponding meson; $\theta_{\rho} = \theta_{\omega} = 81.8^{\circ}, \theta_{\rho}^{0} = \theta_{\omega}^{0} = 61.5^{\circ}, \theta_{\phi} = 68.4^{\circ}$ and $\theta_{\phi}^{0} = 57.13^{\circ}$ are the mixing angles \cite{Volkov:2017arr}. 
       
For the $\eta$ mesons, the factor $A$ takes a slightly different form. This is due to the fact that 
in the case of the $\eta$ meson four states are mixed \cite{Volkov:2022jfr}:
        \begin{eqnarray}
            A^{u}_{M} & = & g_{\eta^{u}} a^{u}_{1M} + g'_{\eta^{u}} a^{u}_{2M} f_{uu}(k_{\perp}^{2}), \nonumber\\
            A^{s}_{M} & = & g_{\eta^{s}} a^{s}_{1M} + g'_{\eta^{s}} a^{s}_{2M} f_{ss}(k_{\perp}^{2}).
        \end{eqnarray}
        
Here $f\left(k_{\perp}^{2}\right) = \left(1 + d k_{\perp}^{2}\right)\Theta(\Lambda^{2} - k_{\perp}^2)$ is 
the form-factor describing the first radially excited meson states. The slope parameters, 
$d_{uu} = -1.784 \times 10^{-6} \textrm{MeV}^{-2}$ and $d_{ss} = -1.737 \times 10^{-6} \textrm{MeV}^{-2}$, 
are unambiguously fixed from the condition of constancy of the quark condensate after the inclusion 
of radially excited states and depends only on the quark composition of the corresponding meson.

The values of the mixing ($a^q_i$) parameters are shown in Table \ref{tab_eta} \cite{Volkov:2022jfr}. The $\eta'$ meson corresponds to the physical state $\eta'(958)$ and the $\hat{\eta}$, $\hat{\eta}'$ mesons correspond to the first radial excitation mesons $\eta$ and $\eta'$.
   
\begin{table}[h!]
\begin{center}
\begin{tabular}{ccccc}
\hline
   & $\eta$ & $\hat{\eta}$ & $\eta'$ & $\hat{\eta}'$ \\
\hline
$a^{u}_{1}$		& 0.71			& 0.62            &-0.32             & 0.56    \\
$a^{u}_{2}$		& 0.11			& -0.87           & -0.48            & -0.54   \\
$a^{s}_{1}$               & 0.62                        & 0.19            & 0.56             & -0.67 \\
$a^{s}_{2}$               & 0.06                       & -0.66           & 0.3               & 0.82 \\
\hline
\end{tabular}
\end{center}
\caption{Mixing parameters of $\eta$ mesons.}
\label{tab_eta}
\end{table}   
    
The quark-meson coupling constants have the form
\begin{eqnarray}
	\label{Couplings}
g_{\rho} =g_{\omega} = \left(\frac{2}{3}I_{20}\right)^{-1/2}, \, g'_{\rho} = g'_{\omega}=\left(\frac{2}{3}I_{20}^{f^{2}}\right)^{-1/2}, 
\, g_{\phi} =  \left(\frac{2}{3}I_{02}\right)^{-1/2}, \, g'_{\phi} =  \left(\frac{2}{3}I_{02}^{f^{2}}\right)^{-1/2}, \nonumber\\
\quad g_{\eta^{u}} = \left(\frac{4}{Z_{\eta^{u}}}I_{20}\right)^{-1/2}, \, g'_{\eta^{u}} =  \left(4 I_{20}^{f^{2}}\right)^{-1/2}, 
\, g_{\eta^{s}} =  \left(\frac{4}{Z_{\eta^{s}}}I_{02}\right)^{-1/2}, \, g'_{\eta^{s}} =  \left(4 I_{02}^{f^{2}}\right)^{-1/2}.
\end{eqnarray}
Here $Z_{\eta^{u}}$ and $Z_{\eta^{s}}$ are additional renormalization constants appearing 
in pseudoscalar and axial-vector transitions.

Integrals appearing in the quark loops are
\begin{eqnarray}
	I_{n_{1}n_{2}}^{f^{m}} =
	-i\frac{N_{c}}{(2\pi)^{4}}\int\frac{f^{m}(k^2_{\perp})}{(m_{u}^{2} - k^2)^{n_{1}}(m_{s}^{2} - k^2)^{n_{2}}}\Theta(\Lambda_{3}^{2} - k^2_{\perp})
	\mathrm{d}^{4}k,
\end{eqnarray}
where $\Lambda_3=1030$ MeV is the cutoff parameter \cite{Volkov:2022jfr}.

\section{Amplitudes and numerical estimations} 
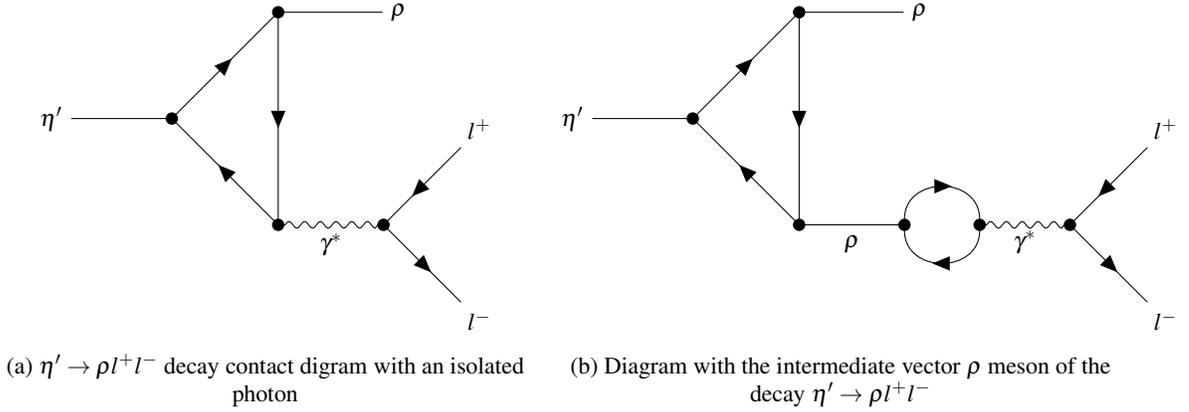
\begin{figure*}[t]
 \centering
  \begin{subfigure}{0.5\textwidth}
   \centering
    \begin{tikzpicture}
     \begin{feynman}
      \vertex (a) {\(\eta' \)};
      \vertex [dot, right=1.6cm of a] (b) {};
      \vertex [dot, above right=2cm of b] (c) {};
      \vertex [dot, below right=2cm of b] (d) {};
      \vertex [right=1.6cm of c] (g) {\(\rho\)};
      \vertex [dot, right=1.4cm of d] (f) {};
      \vertex [above right=1.8cm of f] (e) {\(l^{+}\)};
      \vertex [below right=1.8cm of f] (k) {\(l^{-}\)};
      \diagram* {
        (a) -- [] (b),
        (b) -- [fermion] (c),
        (c) -- [fermion] (d),
        (d) -- [fermion] (b),  
        (c) -- [] (g),         
        (d) -- [photon, inner sep=3pt, edge label'=\(\gamma^{*}\)] (f),
        (f) -- [anti fermion] (e),
        (f) -- [fermion] (k),
      };
     \end{feynman}
    \end{tikzpicture}
   \caption{$\eta' \to \rho l^+l^-$ decay contact digram with an isolated photon}
  \end{subfigure}%
 \centering
 \begin{subfigure}{0.5\textwidth}
  \centering
   \begin{tikzpicture}
    \begin{feynman}
      \vertex (a) {\(\eta' \)};
      \vertex [dot, right=1.6cm of a] (b) {};
      \vertex [dot, above right=2cm of b] (c) {};
      \vertex [dot, below right=2cm of b] (d) {};
      \vertex [right=1.6cm of c] (g) {\(\rho\)};
      \vertex [dot, right=1.4cm of d] (n) {};
      \vertex [dot, right=1.0cm of n] (m) {};     
      \vertex [dot, right=1.2cm of m] (f) {};
      \vertex [above right=1.8cm of f] (e) {\(l^{+}\)};
      \vertex [below right=1.8cm of f] (k) {\(l^{-}\)};
      \diagram* {
        (a) -- [] (b),
        (b) -- [fermion] (c),
        (c) -- [fermion] (d),
        (d) -- [fermion] (b),  
        (c) -- [] (g),         
        (d) -- [inner sep=4pt, edge label'=\(\rho\)] (n),
        (n) -- [fermion, inner sep=1pt, half left] (m),
        (m) -- [fermion, inner sep=1pt, half left] (n), 
        (m) -- [photon, inner sep=2pt, edge label'=\(\gamma^{*}\)] (f),        
        (f) -- [anti fermion] (e),
        (f) -- [fermion] (k),
      };
     \end{feynman}
    \end{tikzpicture}
   \caption{Diagram with the intermediate vector $\rho$ meson of the decay $\eta' \to \rho l^+l^-$}
  \end{subfigure}%
 \caption{Diagrams contributing to the semileptonic decay $\eta' \to \rho l^+l^-$($l=e$ or $\mu$).}
 \label{diagrams}
\end{figure*}

Semileptonic decays of $\rho$, $\omega$, $\phi$, $\eta$ and $\eta'$ mesons are described by the channels with an isolated photon and intermediate neutral vector mesons. The diagrams describing the processes $\eta' \to \rho(\omega)l^+l^-$ are presented in Figure \ref{diagrams}. 

The calculations give the following amplitudes for the $\eta'$ meson decays into a $\rho$ meson and a charged lepton pair:
\begin{eqnarray}
\label{amplitude_1}
\mathcal{M}(\eta' \to \rho l^+l^-) = \frac{\alpha_{em} g_\rho C_{\eta' \rho}}{2 \pi F_\pi} \cdot \frac{1}{s} \left[ 1 + \frac{s}{M_{\rho}^{2} - s - i \sqrt{s}\Gamma_{\rho}} \right] \varepsilon_{\mu\nu\lambda\delta} p_\mu^{\eta'} p_\nu^{\rho} l_\lambda e^{*}_\delta(p^\rho),
\end{eqnarray} 
where $\alpha_{em}=1/137$; $s=(p_{l^+} + p_{l^-})^2$; $C_{\eta' \rho} = 3 \cos(\bar{\theta})$ ($C_{\eta' \omega} = \cos(\bar{\theta})$) and $F_\pi = 92.4$ MeV \cite{Volkov:2022jfr}; $e^{*}_\delta(p^\rho)$ is the polarization vector. The terms in the square brackets in the amplitude (\ref{amplitude_1}) describe the contact diagram and the diagram with the intermediate $\rho$ mesons in the ground state. The values of meson masses and widths are taken from PDG \cite{ParticleDataGroup:2020ssz}. Note that the vector dominance method is automatically reproduced in the NJL model. Vector dominance appears when two channels, in which the considered processes happen, are united. An independent examination of these channels ensures a more detailed study of the processes of interest. 

The decay amplitudes $\rho, \omega, \phi \to \eta l^+ l^-$ and $\phi \to \eta' l^+ l^-$ are obtained by rearranging the ends in the diagram (\ref{diagrams}) and replacing the constants $C_{\rho\eta}=3 \sin(\bar{\theta})$, $C_{\omega\eta}=\sin(\bar{\theta})$, $C_{\phi\eta}=2 \cos(\bar{\theta})F_\pi g_\phi / F_s g_\rho$ and $C_{\phi\eta'}=2 \sin(\bar{\theta})F_\pi g_\phi / F_s g_\rho$.

Using the extended NJL model for the process with mesons in the first radially excited states, we obtain the following amplitude:
\begin{eqnarray}
\label{amplitude_ext}
\mathcal{M}(V' \to \eta_i l^+l^-) = \frac{4 \pi \alpha_{em}}{s} \cdot 4m_u r_V \left[ I_{30}^{V' \eta_i} + I_{30}^{V' V \eta_i} \frac{C_V}{g_V} \frac{s}{M_{V}^{2} - s - i \sqrt{s}\Gamma_{V}} \right] \varepsilon_{\mu\nu\lambda\delta}e^{*}_\mu(p^{V'}) l_\nu p_\lambda^{\eta_i} p_\delta^{V'},
\end{eqnarray} 
where $\eta_i=\eta, \eta'(958), \eta(1295), \eta(1475)$; $r_\rho=1$, $r_\omega=1/3$ and $r_\phi=2m_s/3m_u$; the constants $C_V$ describe the transition of the vector mesons $\rho,\omega, \phi \to \gamma$ through the quark loop
         \begin{eqnarray}
        \label{C_const}
            C_{V}= \frac{1}{\sin{\left(2\theta_{M}^{0}\right)}} \left[\sin{\left(\theta_{M} + \theta_{M}^{0}\right)} + R_{M} \sin{\left(\theta_{M} - \theta_{M}^{0}\right)}\right],
\end{eqnarray}
where $R_{\rho}= R_{\omega}\approx 0.55$, $R_{\phi}\approx 0.43$ \cite{Volkov:2022jfr}.  

The integrals
\begin{eqnarray}
\label{integral}
&& I_{n_1n_2}^{V' V \eta_i...}(m_{u}, m_{s}) = -i\frac{N_{c}}{(2\pi)^{4}} 
 \int\frac{a(k_{\perp}^{2})...b(k_{\perp}^{2})...}{(m_{u}^{2} - k^2)^{n_1}(m_{s}^{2} - k^2)^{n_2}}
\theta(\Lambda_{3}^{2} - k_{\perp}^{2}) \mathrm{d}^{4}k,
\end{eqnarray}
are obtained from the quark triangular loops; $a(k_{\perp}^{2})$ and $b(k_{\perp}^{2})$ are the coefficients
for different mesons defined in (\ref{verteces1}).

The decay amplitudes of the first radially excited mesons $[\eta(1295), \eta(1475)] \to V l^+ l^-$ are obtained by replacing the vertices of the corresponding mesons in the amplitude (\ref{amplitude_ext}). 

The obtained results and the experimentally measured values for the decay widths of the semileptonic decays are given in Table \ref{tab_2}. In this Table comparisons with the results of other theoretical studies are also given.

The precision of the NJL model is determined on the basis of partial conservation of the axial current (PCAC). Based on our previous works, we estimate the uncertainty of the model predictions for relevant widths at the level of 15\% \cite{Volkov:2017arr, Volkov:2022jfr}. 
 
\begin{table}[h!]
\begin{center}
\begin{tabular}{cccccc}
\hline
Decay mode & NJL model & \cite{Branz:2009cd} &  \cite{Hashimoto:1996ny}  &  \cite{Faessler:1999de} &  Data \cite{ParticleDataGroup:2020ssz}  \\
\hline
$\rho \to \eta e^+e^-$		& 0.49			& 0.46            &0.30       & 0.40     & < 1.04  \\
$\rho \to \eta \mu^+\mu^-$		& $2.7\times10^{-5}$	  & --           &$1.12\times10^{-5}$       &   $1.04\times10^{-5}$      & --  \\
$\omega \to \eta e^+e^-$               & 0.059                   & 0.056       & 0.030   &  0.052        & < 0.093  \\
$\omega \to \eta \mu^+\mu^-$        & $9.54\times 10^{-6}$      & --           &$8.50\times 10^{-6}$    &   $15.62\times 10^{-6}$       & --  \\
$\phi \to \eta e^+e^-$	          	& 0.50			& 0.44            &0.48     &    0.46        & $0.459\pm0.017$   \\
$\phi \to \eta \mu^+\mu^-$		& 0.020		& 0.022           &0.023      &     0.028     & < 0.04  \\
$\phi \to \eta' e^+e^-$		& $3.98\times 10^{-3}$	 & $2.1\times 10^{-3}$            &$2.62\times 10^{-3}$  & --     & --   \\
$\eta' \to \rho e^+e^-$		& 0.44			& --            &0.341     &   --    & $0.45\pm0.22$   \\
$\eta' \to \omega e^+e^-$		& 0.042			& --            &0.024     &   0.037     & $0.037\pm 0.007$  \\
$\eta \to \gamma e^+e^-$		& $10.3\times 10^{-3}$		& $8.5\times 10^{-3}$     &$7.78\times 10^{-3}$     &  $8.51\times 10^{-3}$   & $(9.03 \pm 0.52)\times 10^{-3}$  \\
$\eta \to \gamma \mu^+\mu^-$		& $0.349\times 10^{-3}$			& $0.4\times 10^{-3}$            &$0.37\times 10^{-3}$  &    $0.39\times 10^{-3}$   & $(0.40 \pm 0.05)\times 10^{-3}$  \\
$\eta' \to \gamma e^+e^-$		& $90.8\times 10^{-3}$			& $90\times 10^{-3}$            &$85\times 10^{-3}$    &  $78.9\times 10^{-3}$      & $(92.3 \pm 5.07)\times 10^{-3}$  \\
$\eta' \to \gamma \mu^+\mu^-$		& $15\times 10^{-3}$			& $20\times 10^{-3}$            &$18.3\times 10^{-3}$   &   $15.2\times 10^{-3}$  & $(21.2 \pm 5.26)\times 10^{-3}$  \\
\hline
\end{tabular}
\end{center}
\caption{Semileptonic decay widths in keV.}
\label{tab_2}
\end{table}  

\begin{table}[h!]
\begin{center}
\begin{tabular}{ccc}
\hline
 Decay mode  & $\Gamma_{e^+e^-}$, keV & $\Gamma_{\mu^+ \mu^-}$, keV  \\
\hline
$\rho' \to \eta l^+ l^-$		& 2.22			& 0.84 \\
$\omega' \to \eta l^+ l^-$		& 0.65			& 0.49 \\
$\phi' \to \eta l^+ l^-$		& 5.30			& 4.60 \\

$\rho' \to \eta'(957) l^+ l^-$      & 2.00			& 0.19 \\
$\omega' \to \eta'(957) l^+ l^-$		& 0.071			& $1.31\times 10^{-3}$ \\
$\phi' \to \eta'(957) l^+ l^-$		& 1.33			& 0.12 \\

$\rho' \to \eta(1295) l^+ l^-$		& 0.038			&$7.4\times 10^{-5}$ \\
$\omega' \to \eta(1295) l^+ l^-$		& $1.28\times 10^{-3}$			& $8.38\times 10^{-6}$ \\
$\phi' \to \eta(1295) l^+ l^-$		& 0.24			& $4.45\times 10^{-3}$ \\

$\phi' \to \eta(1475) l^+ l^-$		& 0.037			& $1.59\times 10^{-7}$ \\

$\eta(1295) \to \rho l^+ l^-$		& 19.02			& 1.07 \\
$\eta(1295) \to \omega l^+ l^-$		& 2.03			& 0.10 \\
$\eta(1295) \to \phi l^+ l^-$		& 1.33			& $2.73\times 10^{-3}$ \\

$\eta(1475) \to \rho l^+ l^-$		& 0.015			& $0.548\times 10^{-3}$ \\
$\eta(1475) \to \omega l^+ l^-$		& $1.66\times 10^{-3}$			& $5.73\times 10^{-5}$ \\
$\eta(1475) \to \phi l^+ l^-$		& 0.11			& $3.31\times 10^{-3}$ \\
\hline
\end{tabular}
\end{center}
\caption{Predictions of the NJL model for the semileptonic decay widths of vector and eta mesons in the first radially excited states. }
\label{tab_3}
\end{table} 

\section{Conclusions}
As noted in the Introduction, experiments are currently being actively carried out to study the $\eta$ meson physics. At the same time, high statistics are expected in future planned experiments. Consideration of the production of a charged lepton pair instead of a photon in the final state helps to get rid of the background influence in the experimental measurement of the decay widths. 

Our calculations are close to theoretical works in which the transition form factor method is used to describe semileptonic decays \cite{Terschlusen:2010gtc, Schneider:2012ez}. However, the approach based on the transition form factor requires the use of many additional arbitrary parameters. In the NJL model, all the necessary parameters are fixed in advance \cite{Volkov:1986zb, Volkov:2017arr}. Moreover, in the decays considered here, all processes proceed through quark loops of the anomalous type which do not contain ultraviolet divergences. This also contributes to obtaining obvious final results. 

Semileptonic decays of vector mesons involving $\eta$ and $\eta'$ mesons in the ground state have been fairly well studied from both experimental and theoretical points of view. At the same time, few experimental studies with the participation of mesons in radially excited states have been carried out \cite{BESIII:2018dim}. Therefore, it is of particular interest to give a series of predictions for semileptonic decays of radially excited mesons, which may be useful in carrying out the corresponding experimental studies. 
 
\subsection*{Acknowledgments}
We are grateful to A.B. Arbuzov for his interest in our work and important remarks which 
improved the paper; This research has been funded by the Science Committee of the Ministry 
of Education and Science of the Republic of Kazakhstan (Grant No. AP13067963).

\end{document}